\documentclass[prd,twocolumn,floatfix,amsmath,nofootinbib,amssymb,floatfix]{revtex4}
\usepackage{graphicx,color,dcolumn,booktabs,bm,multirow}
\usepackage{longtable,lscape}
\usepackage{txfonts}
\usepackage{overpic}
\usepackage{amssymb}
\usepackage{array}
\usepackage{adjustbox}
\usepackage{indentfirst}
\usepackage{feynmf}   
\usepackage{slashed}  
\usepackage{cases}
\usepackage{booktabs}
\usepackage{multirow}
\usepackage{color}
\usepackage{epstopdf}
\usepackage{multirow,array}
\usepackage{tensor}
\usepackage{tabularx}
\usepackage{booktabs}
\usepackage[colorlinks,
            citecolor=blue,
            anchorcolor=red,
            menucolor=red,
            linkcolor=red,
            filecolor=red,
            runcolor=red,
            urlcolor=blue,
            frenchlinks=red]{hyperref}

\begin{document}

\title{Long-Distance Contributions to Light Vector Mesons and Radiative Decays of  $\eta_{c}$} 
\author{Xiao-Yu Qi$^{1}$}
\author{Qi Wu$^{2}$}
\author{Dian-Yong Chen$^{1,3}$\footnote{Corresponding author}}\email{chendy@seu.edu.cn}
\affiliation{
$^1$ School of Physics, Southeast University, Nanjing 210094, China\\
$^2$ \mbox{School of Physics, Henan Normal University, Xinxiang 453007, China}\\
$^3$ \mbox{Lanzhou Center for Theoretical Physics, Lanzhou University, Lanzhou 730000, China}\\
}
\date{\today}

\begin{abstract}
In the present work, we investigate the long-distance contributions to light hadrons and radiative decays of $\eta_c$, where the long-distance contribution is phenomenologically described by the meson loops. Our estimations indicate that the branching fractions of $\eta_c \to \mathcal{VV},\  \mathcal{V}=(\rho,\omega, \phi)$ can be roughly reproduced in the range $1.4<\alpha <1.6$. While the parameter range for $\eta_c \to \gamma \mathcal{V}$ is determined by the similarity of $\eta_c \to \gamma \mathcal{V}/\mathcal{VV}$ and $\chi_{c1} \to \gamma \mathcal{V}/\mathcal{VV}$. In this parameter range, our estimations indicate that the branching fractions for the processes $\eta_c \to \gamma \rho^0$ and $\eta_c \to \gamma \omega$ are of order $10^{-4}$, while the one for the process $\eta_c \to \gamma \phi$ is of order $10^{-5}$. The estimations in the present work are expected to be accessible by future experiments at BESIII and Belle II.
\end{abstract}

\maketitle

\section{Introduction}
\label{Sec:introduction}
Charmonium, the bound state of a charm quark and its antiquark, provides a unique laboratory for studying Quantum Chromodynamics (QCD) in the transition region between perturbative and nonperturbative regimes. Since the discovery of the $J/\psi$ in 1974, the charmonium spectrum has been extensively studied and is well described by various quark models with linear confinement potentials, including the nonrelativistic Cornell model~\cite{Eichten:1974af,Eichten:1978tg}, the relativized Godfrey-Isgur model~\cite{Godfrey:1985xj,Barnes:2005pb}, and relativistic quark models~\cite{Ebert:2011jc}. Among the charmonium states, the $\eta_c(1S)$ (hereafter simply $\eta_c$) is the lowest-lying $S$-wave spin-singlet pseudoscalar state and the partner of the $J/\psi$. First observed by the Crystal Ball and Mark II Collaborations~\cite{Partridge:1980vk,Himel:1980dj} with a mass of $(2984\pm4)~\mathrm{MeV}/c^2$, its spin-parity $J^{PC}=0^{-+}$ was subsequently confirmed by the Mark III Collaboration through the sequential decay $J/\psi\to\gamma\eta_c$, $\eta_c\to\phi\phi$~\cite{MARK-III:1984mrk}.

Despite over four decades of study, the properties of the $\eta_c$ remain less well constrained than those of its vector counterpart. The Particle Data Group (PDG) average gives a mass of $2984.1\pm0.4~\mathrm{MeV}$ and a total width of $\Gamma=30.5\pm0.5~\mathrm{MeV}$~\cite{ParticleDataGroup:2024cfk}. The measured mass and width, however, exhibit considerable variation across different production mechanisms, including $p\bar{p}$ annihilation~\cite{FermilabE835:2003ula}, two-photon fusion~\cite{BaBar:2003osb,CLEO:2003gwz,BaBar:2011gos}, $B$ meson decays~\cite{Belle:2002jvr,Belle:2011mee}, and $M_1$ radiative transitions $\psi(1S,2S)\to\gamma\eta_c$~\cite{BES:2003vit,CLEO:2008pln,BESIII:2011ab}. In addition, the exclusive decay modes that have been measured account for only about $60\%$ of the total width, and most measurements suffer from large uncertainties~\cite{ParticleDataGroup:2024cfk}. A more complete understanding of $\eta_c$ decay dynamics therefore remains an important open problem.

Among the hadronic decay channels of the $\eta_c$, the decays into a pair of light vector mesons, $\eta_c\to\mathcal{VV}$ with $\mathcal{V}=(\rho^0,\omega,\phi)$, are of particular interest. These processes are doubly suppressed: they violate the Okubo-Zweig-Iizuka (OZI) rule and are also forbidden by the helicity selection rule (HSR)~\cite{Brodsky:1981kj,Chernyak:1981zz,Chernyak:1983ej}. Perturbative QCD (pQCD) therefore predicts very small branching fractions. However, measurements by the BESIII~\cite{BESIII:2019hek,BESIII:2016tqi,BESIII:2025moh} and Belle~\cite{Belle:2012qqr} Collaborations reveal branching fractions that are orders of magnitude larger than these pQCD expectations, indicating that nonperturbative effects play a dominant role. The hadronic loop mechanism, in which the $\eta_c$ dissociates into intermediate charmed mesons that subsequently rescatter into light vector mesons, provides a natural evasion of both the OZI and HSR suppressions~\cite{Zhao:2006cx,Wang:2010iq,Wang:2012wj,Chen:2010re,Chen:2009ah}. The meson loop mechanism can be considered as the phenomenological description of the long-distance contributions. This mechanism has also been successfully applied to $\chi_{c1}$ decays, where the HSR-violating decays $\chi_{c1}\to\mathcal{VV}$ are observed with branching fractions comparable to those of the helicity-allowed $\chi_{c0/2}$ channels~\cite{BESII:2011hcd,Liu:2009vv,Chen:2009ah}. 

\begin{table*}[htb]
\renewcommand{\arraystretch}{1.5}
\centering
\caption{PDG branching fractions for $\eta_c\to\mathcal{VV}$, $\chi_{c1}\to\mathcal{VV}$, and $\chi_{c1}\to\gamma\mathcal{V}$~\cite{ParticleDataGroup:2024cfk}.}
\label{Table:brPDG}
\begin{tabular}{p{1.5cm}<\centering p{2.2 cm}<\centering p{2.2cm}<\centering p{2.6cm}<\centering p{2.6 cm}<\centering p{2.2cm}<\centering p{2.5cm}<\centering}
\toprule[0.75pt]\toprule[0.75pt]
\multirow{2}{*}{State} & \multicolumn{3}{c}{$\mathcal{VV}$} & \multicolumn{3}{c}{$\gamma\mathcal{V}$} \\
 & $\rho\rho$ & $\omega\omega$ & $\phi\phi$ & $\gamma\rho$ & $\gamma\omega$ & $\gamma\phi$ \\
\midrule[0.75 pt]
$\eta_c$
 & $(1.8\pm0.4)\%$ & $(2.7\pm0.9)\times10^{-3}$ & $(1.8\pm0.4)\times10^{-3}$
 & -- & -- & -- \\[2ex]
$\chi_{c1}$
 & -- & $(5.7\pm0.7)\times10^{-4}$ & $(4.26\pm0.21)\times10^{-4}$
 & $(2.16\pm0.17)\times10^{-4}$ & $(6.8\pm0.8)\times10^{-5}$ & $(2.4\pm0.5)\times10^{-5}$ \\[2ex]
\bottomrule[0.75pt]\bottomrule[0.75pt]
\end{tabular}
\end{table*}

A related puzzle concerns the radiative decays $\chi_{c1}\to\gamma\mathcal{V}$. The pQCD estimates of Ref.~\cite{Gao:2006bc} yield branching fractions of order $10^{-6}$--$10^{-5}$, which are at least an order of magnitude below the experimental values reported by CLEO and BESIII~\cite{CLEO:2008sah,BESIII:2011ysp}. In Ref.~\cite{Chen:2010re}, it was demonstrated that including the hadronic loop contribution brings the theoretical predictions for $\chi_{c1}\to\gamma\mathcal{V}$ into agreement with the CLEO data. These findings underscore the importance of nonperturbative dynamics in charmonium decays. The current PDG branching fractions for $\eta_c\to\mathcal{VV}$, $\chi_{c1}\to\mathcal{VV}$, and $\chi_{c1}\to\gamma\mathcal{V}$ are summarized in Table~\ref{Table:brPDG}. In the present work, we extend the hadronic loop mechanism to the radiative decays $\eta_c\to\gamma\mathcal{V}$ ($\mathcal{V}=\rho^0,\omega,\phi$). In the framework of nonrelativistic QCD (NRQCD), these decays are estimated to be of order $10^{-7}$--$10^{-6}$~\cite{Gao:2007fv}. Our goal is to determine whether nonperturbative effects, as encoded in the hadronic loop mechanism, can significantly enhance these rates and bring them within reach of current experiments. For example, the BESIII Collaboration, with its large $J/\psi$ data sample, is particularly well suited for studying $\eta_c$ radiative decays via $J/\psi\to\gamma\eta_c$~\cite{BES:2005bba,BESIII:2016tqi,BESIII:2019hek}.

This paper is organized as follows. In Sec.~\ref{sec:Sec2}, we introduce the effective Lagrangian approach and the hadronic loop mechanism, and derive the amplitudes for $\eta_c\to\mathcal{VV}$ and $\eta_c\to\gamma\mathcal{V}$. Section~\ref{Sec:Num} presents the numerical results and relevant discussions, including the determination of the model parameter from $\eta_c\to\mathcal{VV}$ data and the predictions for $\eta_c\to\gamma\mathcal{V}$, and the last section is devoted to a short summary.

\section{The light meson strong decays and radiative decays of the $\eta_{c}$}
\label{sec:Sec2}

\begin{figure}[t]
	\begin{tabular}{ccc}
		\centering
		\includegraphics[width=2.8cm]{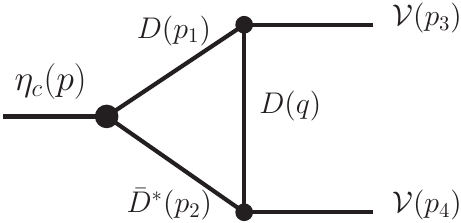}&
		\includegraphics[width=2.8cm]{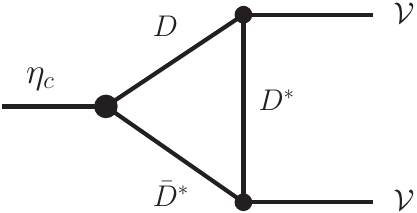}&
		\includegraphics[width=2.8cm]{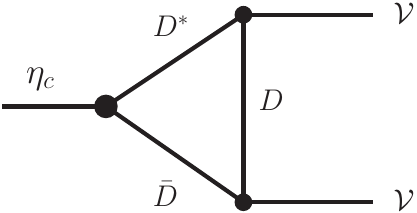}\\
		\\
		$(A_1)$ & $(A_2)$ & $(A_3)$ \\
		\\
		\includegraphics[width=2.8cm]{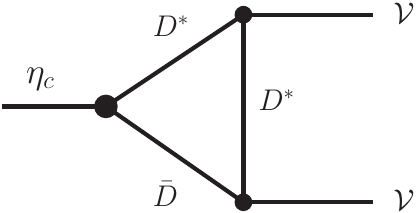}&
		\includegraphics[width=2.8cm]{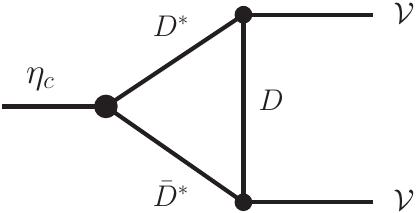}&
		\includegraphics[width=2.8cm]{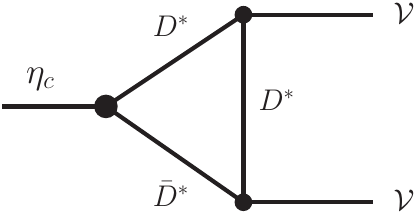}\\
		\\
		$(A_4)$ & $(A_5)$ & $(A_6)$ \\
	\end{tabular}
	\caption{Diagrams contributing to $\eta_{c}\to \mathcal{V}\mathcal{V}$ at the hadron level, where the momenta of the involved particles are labeled in diagram ($A_1$).
		\label{Fig:Tri1}}
\end{figure}

\begin{figure}[t]
	\begin{tabular}{ccc}
		\centering
		\includegraphics[width=2.8cm]{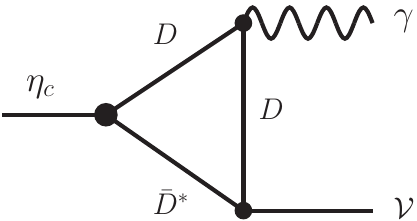}&
		\includegraphics[width=2.8cm]{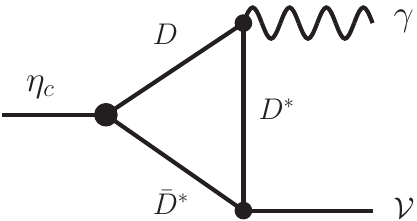}&
		\includegraphics[width=2.8cm]{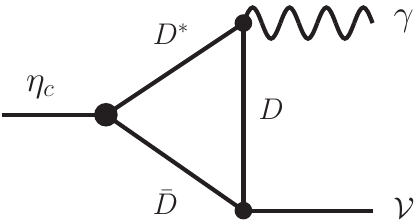}\\
		\\
		$(B_1)$ & $(B_2)$ & $(B_3)$ \\
		\\
		\includegraphics[width=2.8cm]{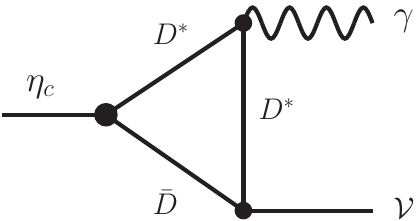}&
		\includegraphics[width=2.8cm]{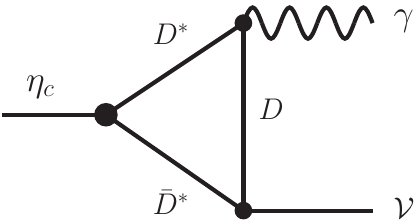}&
		\includegraphics[width=2.8cm]{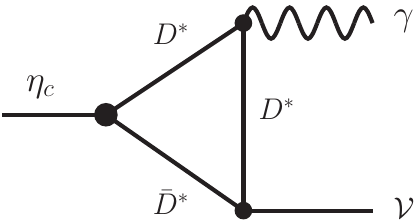}\\
		\\
		$(B_4)$ & $(B_5)$ & $(B_6)$ \\
	\end{tabular}
	\caption{The same as Fig.~\ref{Fig:Tri1} but for $\eta_{c}\to\gamma \mathcal{V}$. 
		\label{Fig:Tri2}}
\end{figure}

As shown in Table~\ref{Table:brPDG}, the branching fractions of $\eta_c\to\mathcal{V}\mathcal{V}$ have been reported. Thus, we can take the measured $\eta_c \to \mathcal{VV}$ as a scale to investigate the meson loop contributions to the decay processes $\eta_c \to \gamma \mathcal{V}$. At the hadron level, the long-distance contribution to the decays $\eta_c \to \mathcal{VV}$ and $\eta_c \to \gamma\mathcal{V}$ can be described in a similar manner. In this framework, the $\eta_c$ first dissociates into two virtual charmed mesons, which subsequently interact via the exchange of another appropriate charmed meson. Consequently, charmed-meson hadronic loops act as a bridge connecting the initial and final states, as illustrated in Figs.~\ref{Fig:Tri1}-\ref{Fig:Tri2}. This hadronic loop mechanism thus offers a tractable approach for computing non-perturbative, long-distance effects in charmonium decays. A more comprehensive review of the meson loop contributions in charmonium decays is presented in the recent review~\cite{Bai:2026atm}.

\subsection{Effective Lagrangians}

In the present estimations, the meson loops are evaluated at the hadron level, and the interaction between the involved hadrons is depicted by effective Lagrangians, which are constructed based on the heavy quark limit and the chiral symmetry. In the heavy quark limit, the $S$-wave charmed and anti-charmed mesons degenerate as~\cite{Lipkin:1986av,Casalbuoni:1996pg,Kaymakcalan:1983qq,Oh:2000qr},
\begin{eqnarray}
H_1&=&\frac{1+v\!\!\!\slash}{2} \Big[ \mathcal{D}^{\ast\mu}\gamma_\mu- \mathcal{D}\gamma_5 \Big] ,\nonumber\\
H_2&=&\Big[\bar{\mathcal{D}}^{\ast\mu}\gamma_\mu-\bar{\mathcal{D}}\gamma_5\Big]\frac{1-v\!\!\!\slash}{2},\nonumber\\
\bar{H}_{1,2}&=&\gamma^0 H^\dagger_{1,2}\gamma^0,
\end{eqnarray}
respectively, with $\mathcal{D}^{(\ast)}=\left(D^{(\ast)0}, D^{(\ast) +}, D_s^{(\ast) +}\right)$.

Unlike the heavy-light meson, the heavy quark flavor symmetry does not hold anymore for heavy quarkonium since the kinetic energy operator includes a factor of $1/m_Q$. However, the degeneracy is still expected under the rotations of the two heavy quark spins, thus one can build heavy quarkonium multiplets for each value of relative angular momentum $\ell$. The $S-$wave charmonia with $\ell =0$ could be expressed by the $4\times 4$ matrix, which reads~\cite{Mannel:1997ky, Brambilla:2004jw, Colangelo:2003sa},
\begin{eqnarray}
	\mathcal{R}&=&\frac{1+v\!\!\!\slash}{2}\left[\psi^{\mu}\gamma_\mu-\eta_{c}\gamma_5\right]\frac{1-v\!\!\!\slash}{2}.
\end{eqnarray}

The general Lagrangians describing the coupling of $S-$ wave charmonia to a pair of $S-$ wave charmed and anti-charmed mesons read,
\begin{eqnarray}
	\mathcal{L_S}=ig_1 \mathrm{Tr}\Big[\mathcal{R}\bar{H}_2 {\stackrel{\leftrightarrow}{\partial_\mu}}\gamma_\lambda \bar{H}_1\Big]+\mathrm{H.c.}.
	\label{Eq:LagT}
\end{eqnarray}
With the above effective Lagrangian, we can obtain the specific form of the effective interactions relevant to the current calculations, denoted as,
\begin{eqnarray}
\mathcal{L}_{\eta_c\mathcal{D}^\ast\mathcal{D}^{(\ast)}}&=&-ig_{\eta_c \mathcal{D}^\ast\mathcal{D}}\eta_c \Big(\mathcal{D}\stackrel{\leftrightarrow}{\partial_\mu}\mathcal{D}^{\ast\dagger\mu}+\mathcal{D}^{\ast\mu}\stackrel{\leftrightarrow}{\partial_\mu}\mathcal{D}^\dagger\Big)\nonumber\\
&&-g_{\eta_c \mathcal{D}^\ast\mathcal{D}^\ast}\varepsilon^{\mu\nu\alpha\beta}\partial_{\mu}\eta\mathcal{D}^{\ast}_{\nu}\stackrel{\leftrightarrow}{\partial_\alpha}\mathcal{D}^{\ast\dagger}_\beta .
\end{eqnarray}

Taking into account the chiral symmetry and the heavy quark limit, the effective interactions between light vector mesons and a pair of charm mesons are constructed as~\cite{Falk:1992cx, Yan:1992gz, Cheng:1992xi, Wise:1992hn},
\begin{eqnarray}
\mathcal{L}_{\mathcal{D}^{(*)} \mathcal{D}^{(*)}\mathcal{V}}&=&-i g_{\mathcal{D} \mathcal{D} \mathcal{V}} \mathcal{D}_{i}^{\dagger} \stackrel{\leftrightarrow}\partial_{\mu} \mathcal{D}^{j}\big(\mathcal{V}^{\mu}\big)_{j}^{i}-2 f_{\mathcal{D}^{*} \mathcal{D} \mathcal{V}} \varepsilon^{\mu \nu \alpha \beta}\nonumber \\
&& \times \big(\partial_{\mu} \mathcal{V}_{\nu}\big)_{j}^{i}\big(\mathcal{D}_{i}^{\dagger}\stackrel{\leftrightarrow} \partial_{\alpha} \mathcal{D}_{\beta}^{* j}-\mathcal{D}_{\beta i}^{* \dagger} \stackrel{\leftrightarrow}{\partial}_{\alpha} \mathcal{D}^{j}\big)\nonumber \\
&&+i g_{\mathcal{D}^{*} \mathcal{D}^{*} \mathcal{V}} \mathcal{D}_{i}^{* \dagger \nu} \stackrel{\leftrightarrow}{\partial}_{\mu} \mathcal{D}_{\nu}^{* j}\big(\mathcal{V}^{\mu}\big)_{j}^{i}\nonumber \\
&&+4 i f_{\mathcal{D}^{*} \mathcal{D}^{*} \mathcal{V}} \mathcal{D}_{i}^{* \dagger \mu}\big(\partial_{\mu} \mathcal{V}^{\nu}-\partial^{\nu} \mathcal{V}_{\mu}\big)_{j}^{i} \mathcal{D}_{\nu}^{* j},
\end{eqnarray}
where $\mathcal{V}$ is the matrix form of the light vector mesons, and their concrete forms are~\cite{Casalbuoni:1996pg,Casalbuoni:1992gi,Casalbuoni:1992dx},
\begin{eqnarray}
\mathcal{V} &=& \left(\begin{array}{ccc}
\frac{1}{\sqrt{2}}\left(\rho^{0}+\omega\right) & \rho^{+} & K^{*+} \\
\rho^{-} & \frac{1}{\sqrt{2}}\left(-\rho^{0}+\omega\right) & K^{* 0} \\
K^{*-} & \bar{K}^{* 0} & \phi
\end{array}\right).
\end{eqnarray}

The effective Lagrangians generating the electromagnetic interaction vertices are~\cite{Dong:2009uf,Chen:2010re},
\begin{eqnarray}
\mathcal{L}_{DD\gamma} &=& ieA_{\mu}D^{\dagger}\stackrel{\leftrightarrow}{\partial}^{\mu}D, \nonumber\\
\mathcal{L}_{D^{*}D^{*}\gamma} &=& ieA_{\mu}\left\{g^{\alpha\beta}D_{\alpha}^{*\dagger}\stackrel{\leftrightarrow}{\partial}^{\mu}D_{\beta}^{*} + g^{\mu\beta}D_{\alpha}^{*\dagger}\partial^{\alpha}D_{\beta}^{*} - g^{\mu\alpha}\partial^{\beta}D_{\alpha}^{*\dagger}D_{\beta}^{*}\right\},\nonumber\\
\mathcal{L}_{D^{*} D \gamma}&=&\frac{e}{4} g_{D^{*} D \gamma} \varepsilon^{\mu \nu \alpha \beta} F_{\mu \nu} D_{\alpha \beta}^{*} D^{\dagger},
\end{eqnarray}
where $F_{\mu\nu} = \partial_\mu A_\nu - \partial_\nu A_\mu$ and $D^{\ast}_{\mu\nu} = \partial_\mu D^{\ast}_\nu - \partial_\nu D^{\ast}_\mu$ is the stress tensor of the photon and vector charmed meson, respectively.

\subsection{Amplitudes for $\eta_c\to \mathcal{VV}$}

Here we take the following picture to describe $\eta_c \to \mathcal{V}\mathcal{V}$. $\eta_c$ could couple to $\mathcal{D}\mathcal{D}^{\ast}$ / $\mathcal{D}^{\ast}\mathcal{D}^{\ast}$, and this charmed meson pair transits into light vector meson pairs by exchanging a proper charmed meson. The sketch diagrams contributing to $\eta_c \to \mathcal{V}\mathcal{V}$ are listed in Fig.~\ref{Fig:Tri1} and the corresponding amplitudes constructed by effective Lagrangians read,
\begin{widetext}
\begin{eqnarray}
\mathcal{M}_{A_1}&=&i^3 \int\frac{d^4 q}{(2\pi)^4}
	\Big[g_{\eta_{c} \mathcal{D}^{*} \mathcal{D}}\left(p_{2\mu}- p_{1\mu}\right)\Big] \left[-g_{ \mathcal{D}\mathcal{D}\mathcal{V}} \epsilon^{\lambda}\left(p_{3}\right) \left(p_{1\lambda}+q_{\lambda}\right)\right]\Big[2f_{\mathcal{D}^*\mathcal{D}\mathcal{V}}\varepsilon_{\kappa\tau\phi\xi}p_4^\kappa\epsilon^\tau(p_4)\big(q^\phi-p_2^\phi\big)\Big] \nonumber\\
	& \times &\frac{1}{p^2_1-m^2_1} \frac{-g^{\alpha\xi}+p_2^{\alpha} p_2^{\xi} /m^2_2}{p^2_2-m^2_2}\frac{1}{q^2-m^2_q}\mathcal{F}^2(m_q^2,q^2),\nonumber\\
%
\mathcal{M}_{A_2}&=&i^3 \int\frac{d^4 q}{(2\pi)^4}
	\Big[g_{\eta_{c} \mathcal{D}^{*} \mathcal{D}} \left(p_{2\mu}- p_{1\mu}\right)\Big]\left[2f_{\mathcal{D}^*\mathcal{D}\mathcal{V}}\varepsilon_{\chi\lambda\sigma\rho}p_3^\chi\epsilon^\lambda(p_3)\big(q^\sigma+p_1^\sigma\big)\right]\left[g_{\mathcal{D}^{*} \mathcal{D}^{*}\mathcal{V}}g_{\tau \phi} \epsilon^{\tau}\left(p_{4}\right)\left(q^{\xi}-p^{\xi}_{2}\right)\right.\nonumber\\
&-&\left.4f_{\mathcal{D}^{*} \mathcal{D}^{*}\mathcal{V}} \epsilon^{\tau}\left(p_{4}\right)\left(p_4^{\xi}g_{\tau \phi}-p^{\phi}_{4}g_{\tau \xi}\right)\right] \frac{1}{p^2_1-m^2_1}\frac{-g^{\alpha\xi}+p_2^{\alpha} p_2^{\xi} /m^2_2}{p^2_2-m^2_2} \frac{-g^{\rho\phi}+q^{\rho} q^{\phi} /m^2_q}{q^2-m^2_q}\mathcal{F}^2(m_q^2,q^2),\nonumber\\
\mathcal{M}_{A_3}&=&i^3 \int\frac{d^4 q}{(2\pi)^4}
	\Big[g_{\eta_{c} \mathcal{D}^{*} \mathcal{D}} \left(p_{2\mu}- p_{1\mu}\right)\Big] \left[-2f_{\mathcal{D}^*\mathcal{D}\mathcal{V}}\varepsilon_{\chi\lambda\sigma\rho}p_3^\chi\epsilon^\lambda(p_3)\big(q^\sigma+p_1^\sigma\big)\right]  \Big[g_{ \mathcal{D}\mathcal{D}\mathcal{V}} \epsilon^{\tau}\left(p_{4}\right) \left(p_{2\tau}-q_{\tau}\right) \Big] \nonumber\\
	&\times&\frac{-g^{\alpha\rho}+p_1^{\alpha} p_1^{\rho} /m^2_1}{p^2_1-m^2_1}\frac{1}{p^2_2-m^2_2} \frac{1}{q^2-m^2_q}\mathcal{F}^2(m_q^2,q^2),\nonumber\\
	\mathcal{M}_{A_4}&=&i^3 \int\frac{d^4 q}{(2\pi)^4}
	\Big[g_{\eta_{c} \mathcal{D}^{*} \mathcal{D}}\left(p_{2\mu}- p_{1\mu}\right)\Big] \left[g_{\mathcal{D}^{*} \mathcal{D}^{*}\mathcal{V}}g_{\lambda \sigma}
	\epsilon^{\lambda}\left(p_{3}\right)\left(q^{\rho}+p^{\rho}_{1}\right)+4f_{\mathcal{D}^{*} \mathcal{D}^{*}\mathcal{V}} \epsilon^{\lambda}\left(p_{4}\right)\left(p_3^{\rho}g_{\lambda \sigma}-p^{\sigma}_{3}g_{\lambda \rho}\right)\right]\nonumber\\
	&\times& \Big[2f_{\mathcal{D}^*\mathcal{D}\mathcal{V}}\varepsilon_{\kappa\tau\phi\xi}p_4^\kappa\epsilon^\tau(p_4)\big(p_2^\phi-q^\phi\big)\Big] \frac{-g^{\alpha\sigma}+p_1^{\alpha} p_1^{\sigma} /m^2_1}{p^2_1-m^2_1} \frac{1}{p^2_2-m^2_2} \frac{-g^{\rho\xi}+q^{\rho} q^{\xi} /m^2_q}{q^2-m^2_q}\mathcal{F}^2(m_q^2,q^2),\nonumber\\
\mathcal{M}_{A_5}&=&i^3 \int\frac{d^4 q}{(2\pi)^4}
	\Big[ig_{\eta_{c2} \mathcal{D}^{*} \mathcal{D}^{*}} \varepsilon_{\alpha\beta\nu\theta}p^{\theta} (p_{2\mu}- p_{1\mu})\Big] \left[-2f_{\mathcal{D}^*\mathcal{D}\mathcal{V}}\varepsilon_{\chi\lambda\sigma\rho}p_3^\chi\epsilon^\lambda(p_3)\big(q^\sigma+p_1^\sigma\big)\right]\Big[2f_{\mathcal{D}^*\mathcal{D}\mathcal{V}}\varepsilon_{\kappa\tau\phi\xi}p_4^\kappa\epsilon^\tau(p_4)\big(q^\phi-p_2^\phi\big)]\nonumber\\
	&&\times\frac{-g^{\alpha\rho}+p_1^{\alpha} p_1^{\rho} /m^2_1}{p^2_1-m^2_1} \frac{-g^{\beta\xi}+p_2^{\beta} p_2^{\xi} /m^2_2}{p^2_2-m^2_2} \frac{1}{q^2-m^2_q}\mathcal{F}^2(m_q^2,q^2),\nonumber\\
\mathcal{M}_{A_6}&=&i^3 \int\frac{d^4 q}{(2\pi)^4}
	\Big[ig_{\eta_{c2} \mathcal{D}^{*} \mathcal{D}^{*}} \varepsilon_{\alpha\beta\nu\theta}p^{\theta}(p_{2\mu}- p_{1\mu})\Big] \left[g_{\mathcal{D}^{*} \mathcal{D}^{*}\mathcal{V}}g_{\lambda \sigma}
	\epsilon^{\lambda}\left(p_{3}\right)\left(q^{\rho}+p^{\rho}_{1}\right)+4f_{\mathcal{D}^{*} \mathcal{D}^{*}\mathcal{V}} \epsilon^{\lambda}\left(p_{4}\right)\left(p_3^{\rho}g_{\lambda \sigma}-p^{\sigma}_{3}g_{\lambda \rho}\right)\right]\nonumber\\
	&\times & \Big[g_{\mathcal{D}^{*} \mathcal{D}^{*}\mathcal{V}}g_{\tau \phi} \epsilon^{\tau}\left(p_{4}\right)\left(q^{\xi}-p^{\xi}_{2}\right) -4f_{\mathcal{D}^{*} \mathcal{D}^{*}\mathcal{V}} \epsilon^{\tau}\left(p_{4}\right)\left(p_4^{\xi}g_{\tau \phi}-p^{\phi}_{4}g_{\tau \xi}\right)\Big] \nonumber\\
	&\times &\frac{-g^{\alpha\sigma}+p_1^{\alpha} p_1^{\sigma} /m^2_1}{p^2_1-m^2_1}\frac{-g^{\beta\xi}+p_2^{\beta} p_2^{\xi} /m^2_2}{p^2_2-m^2_2} \frac{-g^{\rho\phi}+q^{\rho} q^{\phi} /m^2_q}{q^2-m^2_q}\mathcal{F}^2(m_q^2,q^2).
\end{eqnarray}
\end{widetext}

In the above amplitudes, a form factor $\mathcal{F}(m_q^2,q^2)$ is introduced to ensure the convergence of the loop integral and to account for the dynamics and off-shell effects. Generally, various forms of phenomenological form factors, including monopole, dipole, and exponential forms, can be used in the triangle-diagram calculations to regulate high-momentum contributions~\cite{Li:2008ey, Li:2007au,Xiao:2017uve,Gao:2025zhp,Ivanov:2003ge,Yue:2023qgx}. In the present work, we estimate the loop integrals with Feynman parameterization~\cite{Feynman:1949zx} with \texttt{LoopTools} package~\cite{Hahn:1998yk}, for which the pole form is particularly convenient since it can be handled directly in the numerical estimations. We therefore adopt the square of form factor in the pole form in the present estimations, which reads,
\begin{eqnarray}
\mathcal{F}^2\left(m_q^2, q^{2}\right)=\left(\frac{m_q^{2}-\Lambda^{2}}{q^{2}-\Lambda^{2}}\right)^N,
\end{eqnarray}
where the condition $N\ge 2$ must hold to ensure the finiteness of the loop integrals. As indicated in Refs.~\cite{Chen:2009ah,Chen:2010re}, the square of the form factor with $N=4$ is suitable to investigate the light mesons and radiative decays of charmonia. Thus, we take the same form factor in the present work. In addition, the parameter $\Lambda$ in the form factor can be further parameterized as $\Lambda=m_{q}+\alpha \Lambda_{\mathrm{QCD}}$ with $m_q$ to be the mass of the exchanged meson and $\Lambda_{\mathrm{QCD}}=220$ MeV. In general, the model parameter $\alpha$ should be uniform, however its precise value cannot be derived using first principle~\cite{Tornqvist:1993vu,Tornqvist:1993ng,Locher:1993cc,Li:1996yn}. When dealing with specific processes, a more common practice is to determine the exact value of $\alpha$ by comparing experimental measurements with theoretical estimations. It's worth mentioning that for all the involve loops, we have the K\"all\'en function $ \lambda(m_1^2,m_2^2,m_q^2)<0$, so the Landau equation for the leading triangle singularity has no real solution. Therefore, no anomalous threshold appears, and the amplitude admits a single dispersion representation.

The total amplitude for $\eta_c\to \mathcal{VV}$ is,
\begin{eqnarray}
    \mathcal{M}^{\mathrm{Tot}}_{\eta_c \to \mathcal{VV}} =\sum_{i=1}^{6} \mathcal{M}_{A_i}.
\end{eqnarray}
Then, the partial width of $\eta_c\to\mathcal{V}\mathcal{V}$ can be estimated by,
\begin{eqnarray}
&&\Gamma(\eta_c \to \mathcal{V}\mathcal{V})= \frac{1}{8 \pi} \frac{|\vec{p}|}{m_{0}^{2}}\overline{\left|\mathcal{M}_{\eta_c \to\mathcal{V}\mathcal{V}}^{\mathrm{tot}}\right|^{2}},
\end{eqnarray}
where $|\vec{p}|$ is the value of the three momentum of the final state in the $\eta_c$ rest frame.  In the same way, we can construct the amplitudes for $\eta_c \to \gamma\mathcal{V}$, which are collected in Appendix~\ref{Sec:Appetac}.

\section{Numerical Results and discussions}
\label{Sec:Num}

\subsection{Coupling constants}

Before estimating the partial widths of the considered processes, the coupling constants in the effective Lagrangians should be further clarified. In the heavy quark limit, the coupling constants relevant  to $S-$wave charmonia and charmed mesons could be related to a gauge coupling $g_1$ by~\cite{Falk:1992cx,Yan:1992gz,Cheng:1992xi,Wise:1992hn},
\begin{eqnarray}
	g_{\eta_{c} D^{*} D}&=&2 g_{1} \sqrt{m_{\eta_{c}} m_{D} m_{D^{*}}}\ , \nonumber\\		
	g_{\eta_{c} D^{*} D^{*}}&=&2 g_{1} m_{D^{*}} / \sqrt{m_{\eta_{c}}}\ ,
\end{eqnarray}
where the gauge coupling $g_1$ is defined as $g_1= \sqrt{m_{\psi}}/\left(2m_Df_\psi \right)$ with  $f_\psi=416$ MeV being the decay constant of $J/\psi$, which can be estimated from the dilepton decay width of $J/\psi$~\cite{Colangelo:2003sa,ParticleDataGroup:2024cfk}.

Considering the heavy quark limit and chiral symmetry, one can obtain the coupling constants relevant to the light vector mesons, which are~\cite{Falk:1992cx,Chen:2014sra,Yan:1992gz,Cheng:1992xi,Wise:1992hn},
\begin{eqnarray}
g_{\mathcal{D}\mathcal{D}\mathcal{V}}&=&g_{\mathcal{D}^*\mathcal{D}^*\mathcal{V}}=\frac{1}{\sqrt2}\beta_v g_v,\nonumber\\
f_{\mathcal{D}^*\mathcal{D}\mathcal{V}}&=&\frac{f_{\mathcal{D}^*\mathcal{D}^*\mathcal{V}}}{m_\mathcal{D}^*}=\frac{1}{\sqrt2}\lambda_v g_v,
\end{eqnarray}
where $\beta_v=0.9$, $\lambda_v=0.56$, and $g_v={m_\rho}/{f_\pi}$ with $m_\rho$ and  $f_\pi=132\  \mathrm{MeV}$ denoting the mass of $\rho$ meson and decay constant of the pion, respectively.

Calculations based on Lattice QCD ~\cite{Becirevic:2009xp} and QCD sum rules ~\cite{Zhu:1996qy} both predict that the coupling constant for the radiative decay of neutral charmed mesons is positive, while the coupling constant for the radiative decay of charged charmed mesons is negative. In this work, we follow this convention to provide the relevant coupling constants. According to the experimental widths of $D^{\ast+} \to D^{+}\gamma$, the coupling constant $g_{D^{\ast} D\gamma}$ is determined as ~\cite{Dong:2008gb,Chen:2010re,Zhu:1996qy}
\begin{eqnarray}
	g_{D^{*-}D^{+}\gamma} &=& -0.465\ \text{GeV}^{-1}.
\end{eqnarray}
The radiative decay  $D^{\ast0} \to D^{0}\gamma$ belongs to the category of magnetic dipole transitions between the $^3S_1$ and $^1S_0$ states, and these transitions can be successfully described through non-relativistic methods ~\cite{Gasiorowicz:1981jz,Rosner:2013sha}. Based on the updated mass and branching ratio, the estimated value of the radiative decay width (central value) is 21.24 keV ~\cite{Rosner:2013sha}. The coupling constant $g_{D^{*0}D^{0}\gamma}$ is determined accordingly to be,
\begin{eqnarray}
	g_{D^{*0}D^{0}\gamma} &=& 1.85\ \text{GeV}^{-1}.
\end{eqnarray}
Using the results of Lattice QCD calculations for $D_s^{*}\to D_s^{+}\gamma$~\cite{Meng:2024gpd}, we can conclude that
\begin{eqnarray}
	g_{D_s^{*}D_s^{+}\gamma} &=& - 9.90 \times 10^{-2}\ \text{GeV}^{-1}.
\end{eqnarray}

\begin{figure*}[t]
	\centering
	\begin{tabular}{ccc}
		\includegraphics[width=8.4cm]{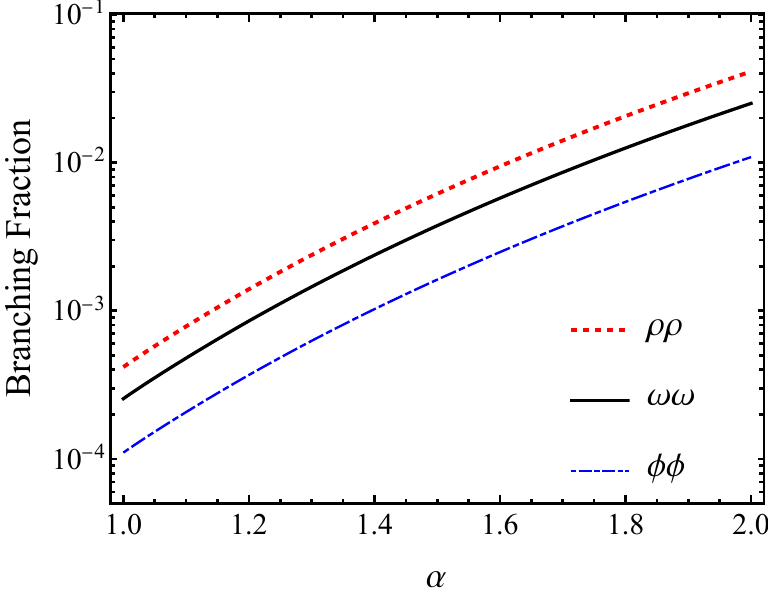}&
		$\hspace{0.5cm}$ &
		\includegraphics[height=6.6cm]{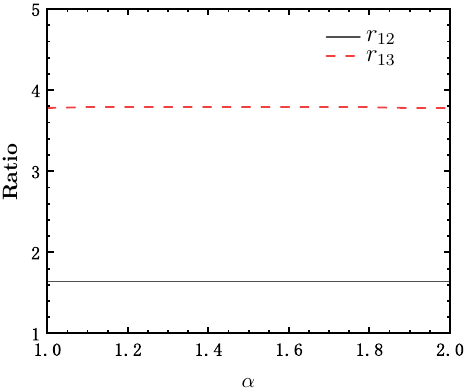}
	\end{tabular}
	\caption{(Color online.) The branching fractions of the light vector meson decays of $\eta_c $ depending on the model parameter $\alpha$(left panel). And the ratios of the branching ratios for  $\eta_c \to \mathcal{VV}$ with $\mathcal{V}=(\rho^0,\omega,\phi)$(right panel). \label{Fig:Br1} }
\end{figure*}

\begin{table}[htb]
\renewcommand{\arraystretch}{1.5}
\caption{Comparison of the present estimations of $\eta_c \to \mathcal{VV}$ and the corresponding experimental measurements~\cite{ParticleDataGroup:2024cfk} \label{Tab:BRVV}}
\begin{tabular}{p{2.5cm}<\centering p{2.5 cm}<\centering p{2.5cm}<\centering}
\toprule[0.8 pt]\toprule[0.8 pt]
Process & Present & PDG~\cite{ParticleDataGroup:2024cfk}\\
\midrule[0.8pt]
$\eta_c\to \rho \rho$ & $(3.9\sim 9.5)\times 10^{-3}$ & $(1.8\pm0.4)\%$ \\
$\eta_c\to \omega \omega$ & $(2.3 \sim 5.7)\times 10^{-3}$ & $(2.7\pm0.9)\times10^{-3}$ \\
$\eta_c\to \phi \phi$ & $(1.0\sim 2.5)\times  10^{-3}$ & $(1.8\pm0.4)\times10^{-3}$\\
\bottomrule[0.8pt]  \bottomrule[0.8pt]
\end{tabular}
\end{table}

\subsection{Branching fractions of $\eta_c\to \mathcal{VV}$}

Besides the above coupling constants, only one parameter, $\alpha$, introduced by the form factor, is undetermined. As has been mentioned in Sec.~\ref{Sec:introduction} and Table~\ref{Table:brPDG}, the processes $\eta_c \to \rho^0\rho^0$, $\eta_c \to \omega \omega$ and $\eta_c \to \phi \phi$ have been measured by the BESIII and Belle Collaborations~\cite{BESIII:2019hek,BESIII:2016tqi,BESIII:2025moh,Belle:2012qqr}, and the PDG average of the branching fractions is $(1.8\pm 0.4)\times 10^{-2}$, $(2.7\pm 0.9)\times 10^{-3}$ and $(1.8\pm 0.4)\times 10^{-3}$, respectively~\cite{ParticleDataGroup:2024cfk}. With the branching fractions of $\eta_c \to \mathcal{VV}$, we can roughly determine the model parameter range by comparing the theoretical estimation in the present work with the experimental data, and within this parameter range, the branching fractions of the discussed processes could be reproduced.

Our estimations of the branching fractions of the light vector meson decays of $\eta_c$ depending on the model parameter $\alpha$ are presented in the left panel of Fig.~\ref{Fig:Br1}. As can be seen from the diagram, the estimated branching fractions increase with the increase of the model parameter $\alpha$. By comparing the present estimations with the PDG average, we find the measured branching fractions can be roughly reproduced in the parameter range $1.4<\alpha <1.6$. Comparisons of the present estimations and experimental measurements are listed in Table~\ref{Tab:BRVV}. From the table, one can find in this parameter range, the branching fractions of the $\omega \omega$ and $\phi \phi$ channels are well consistent with the experimental measurements. As for the $\eta_c \to \rho \rho$ channel, our estimations of the branching fraction are smaller than the measured one, and the measured lower limit is about 1.5 times the estimated upper limit.

In addition, our estimations indicate that the $\alpha$ dependence of the branching fractions are very similar, thus  their ratios are expected to be model independent. Here, we define the following ratios,
\begin{eqnarray}
	r_{12}=\frac{\mathcal{B}(\eta_c\to \rho^0\rho^0)}{\mathcal{B}(\eta_c\to \omega\omega)}, \qquad r_{13}=\frac{\mathcal{B}(\eta_c\to \rho^0\rho^0)}{\mathcal{B}(\eta_c\to \phi\phi)}
	\label{Eq:Ratio1}
\end{eqnarray}	
The $\alpha$ dependence of these ratios is shown in the right panel of Fig.~\ref{Fig:Br1}. One can find these ratios are almost independent on the model parameter $\alpha$, and in the considered parameter range, we have 
\begin{eqnarray}
	r_{12}= 1.64, \qquad r_{13}= 3.78\sim 3.79,\label{eq:r1}
\end{eqnarray}
respectively.

\begin{figure*}[htb]
\centering
\begin{tabular}{ccc}
	\includegraphics[width=8.5cm]{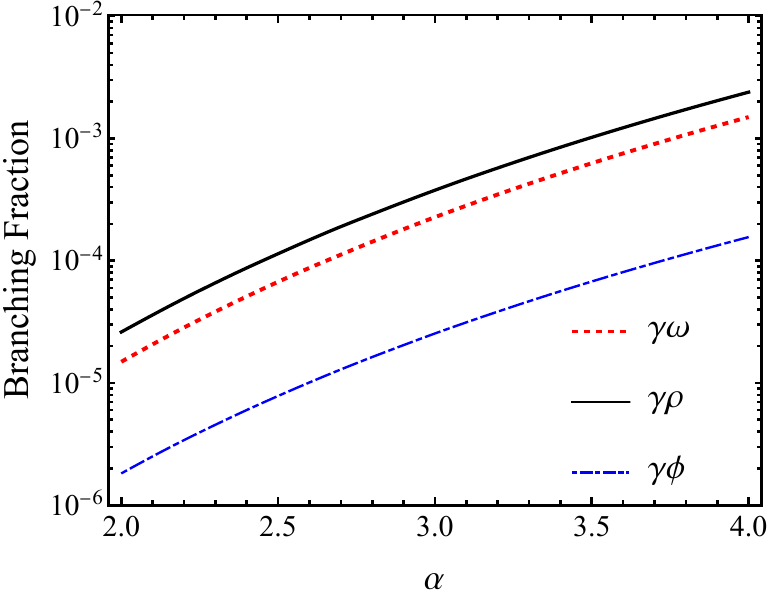}&
	$\hspace{0.5cm}$ &\includegraphics[height=6.6cm]{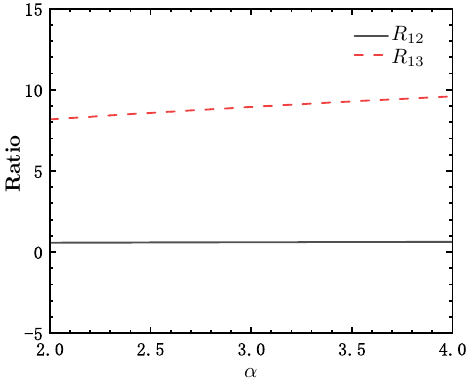}
\end{tabular}
	\caption{(Color online.) The branching ratios of the decays $\eta_c\to \gamma \mathcal{V}$ with $\mathcal{V}=(\rho^0,\omega,\phi)$ depending on the model parameter $\alpha$.(left panel). And the ratios of the branching ratios for  $\eta_c \to \gamma\mathcal{V}$(right panel). \label{Fig:Br2}}
\end{figure*}

\begin{table}[h!]
	\renewcommand\arraystretch{1.5}
	\centering
	\caption{Branching ratios of $\chi_{c1}\to \mathcal{VV}/\gamma\mathcal{V}$ and the corresponding $\alpha$ range by matching the PDG average in Table~\ref{Table:brPDG}~\cite{ParticleDataGroup:2024cfk}.}
	\begin{tabular}{p{2.5cm}<\centering p{3cm}<\centering p{2.5cm}<\centering}
\toprule[0.8 pt] \toprule[0.8 pt]
		Channel ($\chi_{c1} \to$)  &  Branching fractions & $\alpha$ \\
		\midrule[0.8pt]
		$\rho^{0}\rho^{0}$  & - & - \\
		$\omega\omega$ & $(5.0\text{--}6.4) \times 10^{-4}$ & $0.92\text{--}0.96$ \\
		$\phi\phi$  & $(4.05\text{--}4.47) \times 10^{-4}$ & $1.09\text{--}1.12$ \\
		\midrule[0.8pt]
		$\gamma\rho^{0}$  & $(1.99\text{--}2.33) \times 10^{-4}$ & $1.90\text{--}1.95$ \\
		$\gamma\omega$ & $(6.0\text{--}7.6) \times 10^{-5}$ & $1.69\text{--}1.75$ \\
		$\gamma\phi$  & $(1.9\text{--}2.9) \times 10^{-5}$ & $2.42\text{--}2.60$ \\
\bottomrule[0.8pt]  \bottomrule[0.8pt]
	\end{tabular}
	\label{tab:chi_c1_decay}
\end{table}

\subsection{Branching fractions of $\eta_c\to\gamma \mathcal{V}$}

From the above-mentioned decays $\eta_c \to \mathcal{VV}$, we can see that a certain value of $\alpha$ can perfectly match the experimental results. As for the decays $\eta_c \to \gamma\mathcal{V}$ we are concerned with, the two kinds of decay processes should share a similar $\alpha$ range but may not be exactly the same due to the different kinematic space. Since the $\mathcal{VV}$ and $\gamma \mathcal{V}$ decay processes have been experimentally measured and the meson loop contributions have also been studied in Refs.~\cite{Zhao:2006cx,Wang:2010iq,Wang:2012wj,Chen:2010re,Chen:2009ah}, the parameter differences in the $\chi_{c1} \to \mathcal{VV}$ and $\chi_{c1} \to \gamma \mathcal{V}$ processes could serve as a useful reference to investigate the $\eta_c \to \mathcal{VV}$ and $\eta_c \to \gamma \mathcal{V}$ processes. In Ref.~\cite{Chen:2009ah,Chen:2010re}, we have studied $\chi_{c1}\to \mathcal{VV}$ and $\chi_{c1}\to \gamma\mathcal{V}$, and the obtained $\alpha$ range reveals that the $\gamma\mathcal{V}$ channel is larger than that of $\mathcal{VV}$ when matching the experimental data. In the present work, we update these two decay modes by using the same form factor, and the obtained results are collected in Table~\ref{tab:chi_c1_decay}. From the table, one can find for the $\gamma \omega$ process, the $\alpha$ range is greater than that for the $\omega \omega$ process by 0.8, while for the $\gamma \phi$ process, the parameter range is also greater than that for $\phi \phi$, but by $1.3 \sim 1.5$. Considering the similarity of $\chi_{c1} \to \mathcal{VV}/\gamma \mathcal{V}$ and $\eta_c \to \mathcal{VV}/\gamma \mathcal{V}$ and the parameter range determined by $\eta_c \to \mathcal{VV}$, we obtain that the range $2.5<\alpha<3.0$ should be suitable to roughly evaluate the branching fractions of the decay processes $\eta_c\to\gamma \mathcal{V}$.

The $\alpha$ dependence of the branching ratios of $\eta_c \to \gamma \mathcal{V}$ is given in Fig.~\ref{Fig:Br2}. Similar to the case of $\eta_c \to \mathcal{VV}$, the branching fractions of $\eta_c \to \gamma \mathcal{V}$ monotonically increase with the increase of the model parameter. With the range of $2.5<\alpha<3.0$, the branching ratios of $\eta_c\to\gamma \mathcal{V}$ are estimated to be,
\begin{eqnarray}
	BR[\eta_{c}\to \gamma \rho^0]&=&\left(0.67\sim 2.27\right)\times10^{-4},\nonumber\\
	BR[\eta_{c}\to \gamma \omega]&=&\left(1.14\sim 3.77 \right)\times10^{-4},\nonumber\\
	BR[\eta_{c}\to \gamma \phi]&=&\left(0.78\sim 2.54\right)\times10^{-5},\label{eq:Br1}
\end{eqnarray}
which are expected to be tested by future experiments at BESIII or Belle II. Furthermore, the experimental measurements of $\eta_c \to \gamma \mathcal{V}$ would help to constrain the cutoff parameter in charmonium decays via the hadronic loop mechanism and provide a deep insight into nonperturbative QCD effects in the charmonium energy region.

Similar to the case of $\eta_c \to \mathcal{VV}$,  we also estimated the ratios of the branching fractions of considered radiative decay processes, which are defined as,
\begin{eqnarray}
R_{12}=\frac{\mathcal{B}(\eta_c\to\gamma\rho^0)}{\mathcal{B}(\eta_c\to\gamma\omega)}, \qquad R_{13}=\frac{\mathcal{B}(\eta_c\to\gamma\rho^0)}{\mathcal{B}(\eta_c\to\gamma\phi)}.
\label{Eq:Ratio2}
\end{eqnarray}
As shown in the right panel of Fig.~\ref{Fig:Br2}, these ratios are nearly independent of the model parameter $\alpha$ and in the considered parameter range, these ratios are estimated to be,
\begin{eqnarray}
	R_{12}&=& 0.57\sim 0.63,\nonumber\\
	R_{13}&=& 8.17\sim 9.59,\label{eq:r2}
\end{eqnarray}
respectively.  In addition, we also discussed the form factor dependence of these ratios in Appendix~\ref{Sec:Appetac1}.

\section{Summary}
\label{Sec:Summary}

With the world's largest data sample of $J/\psi$ collected by the BESIII detector at the BEPCII electron-positron collider, the decay properties of $\eta_c$ could be measured via the decay process $J/\psi \to \gamma \eta_c$. As the lightest charmonium state, $\eta_c$ can only decay into hadronic final states via $c\bar{c}$ annihilation. Similar to the case of $\chi_{c1} \to \mathcal{VV}$, the meson loop mechanism, which can be considered as a phenomenological description of the long-distance contributions, provides a natural evasion of both the OZI and HSR suppressions for the $\eta_c \to \mathcal{VV}$ processes. The present estimations indicate that in a certain parameter range the measured branching fractions of $\eta_c \to \mathcal{VV}$ can be roughly reproduced.

As for the radiative decay processes, $\eta_c \to \gamma \mathcal{V}$, the branching fractions are predicted to be of order $10^{-7}\sim 10^{-6}$ by the NRQCD estimations. In the present work, we applied the same meson loop mechanism to investigate the processes $\eta_c \to \gamma \mathcal{V}$. We determine the parameter range by the similarity of $\eta_c \to \gamma \mathcal{V}/\mathcal{VV}$ and $\chi_{c1} \to \gamma \mathcal{V}/\mathcal{VV}$. In this parameter range, our estimations indicate that the branching fractions for the processes $\eta_c \to \gamma \rho^0$ and $\eta_c \to \gamma \omega$ are of order $10^{-4}$, while the one for the process $\eta_c \to \gamma \phi$ is of order $10^{-5}$. The estimations in the present work are expected to be accessible by future experiments at BESIII and Belle II.

\section*{ACKNOWLEDGMENTS}
The authors would like to thank Professor Tao Luo for inspiring discussions. This work is supported by the National Natural Science Foundation of China under the Grant Nos. 12175037, 12335001 and 12405093, as well as supported, in part, by National Key Research and Development Program under Grant Nos.2024YFA1610503 and 2024YFA1610504.

\appendix
\section{Amplitudes for $\eta_{c}\to \gamma\mathcal{V}$\label{Sec:Appetac}}
The amplitudes of the process $\eta_{c}(p)\to \gamma(p_3) \mathcal{V}(p_4)$, corresponding to Fig.~\ref{Fig:Tri2}-$(B_1)$-$(B_6)$ where the momentums are the same as that marked in Fig.~\ref{Fig:Tri1}$(A_1)$, reads,

\begin{widetext}
\begin{eqnarray}
	\mathcal{M}_{B_1}&=&i^3 \int\frac{d^4 q}{(2\pi)^4}
	\Big[g_{\eta_{c} \mathcal{D}^{*} \mathcal{D}} \left(p_{2\mu}- p_{1\mu}\right)\Big]\left[e \epsilon_{\gamma}^\lambda\left(p_{3}\right) \left(p_{1\lambda}+q_{\lambda}\right)\right] \Big[2f_{\mathcal{D}^*\mathcal{D}V}\varepsilon_{\kappa\tau\phi\xi}p_4^\kappa\epsilon^\tau(p_4)\big(q^\phi-p_2^\phi\big)\Big]\nonumber\\
	& \times& \frac{1}{p^2_1-m^2_1}  \frac{-g^{\alpha\xi}+p_2^{\alpha} p_2^{\xi} /m^2_2}{p^2_2-m^2_2}\frac{1}{q^2-m^2_q}\mathcal{F}^2(m_q^2,q^2),\nonumber\\
	\mathcal{M}_{B_2}&=&i^3 \int\frac{d^4 q}{(2\pi)^4}
	\Big[g_{\eta_{c} \mathcal{D}^{*} \mathcal{D}}\left(p_{2\mu}- p_{1\mu}\right)\Big]\Big[\frac{eg_{\mathcal{D}^*\mathcal{D}\gamma}}{4}\varepsilon^{\chi\lambda\sigma\rho}\epsilon_{\gamma}^{\eta}(p_3)\big(i p_{3\chi} g_{\lambda\eta}-i p_{3\lambda} g_{\chi\eta}\big)\big(i q_\sigma g_{\rho\theta}-i q_\rho g_{\sigma\theta}\big)\Big]\Big[g_{\mathcal{D}^{*} \mathcal{D}^{*}V}g_{\tau \phi} \epsilon^{\tau}\left(p_{4}\right)\left(q^{\xi}-p^{\xi}_{2}\right)\nonumber\\
	&-&4f_{\mathcal{D}^{*} \mathcal{D}^{*}V} \epsilon^{\tau}\left(p_{4}\right)\left(p_4^{\xi}g_{\tau \phi}-p^{\phi}_{4}g_{\tau \xi}\right)\Big] \frac{1}{p^2_1-m^2_1}\frac{-g^{\alpha\xi}+p_2^{\alpha} p_2^{\xi} /m^2_2}{p^2_2-m^2_2} \frac{-g^{\rho\phi}+q^{\rho} q^{\phi} /m^2_q}{q^2-m^2_q}\mathcal{F}^2(m_q^2,q^2),\nonumber \\
\mathcal{M}_{B_3}&=&i^3 \int\frac{d^4 q}{(2\pi)^4}
	\Big[g_{\eta_{c} \mathcal{D}^{*} \mathcal{D}}\left(p_{2\mu}- p_{1\mu}\right)\Big] \left[\frac{eg_{\mathcal{D}^*\mathcal{D}\gamma}}{4}\varepsilon^{\chi\lambda\sigma\rho}\epsilon_{\gamma}^{\eta}(p_3)\big(i p_{3\chi} g_{\lambda\eta}-i p_{3\lambda} g_{\chi\eta}\big)\big(i p_{1\sigma} g_{\rho\theta}-i p_{1\rho} g_{\sigma\theta}\big)\right]
	   \Big[g_{ \mathcal{D}\mathcal{D}V} \epsilon^{\tau}\left(p_{4}\right) \left(p_{2\tau}-q_{\tau}\right) \Big] \nonumber\\
	&\times& \frac{-g^{\alpha\rho}+p_1^{\alpha} p_1^{\rho} /m^1_2}{p^2_1-m^2_1}\frac{1}{p^2_2-m^2_2} \frac{1}{q^2-m^2_q}\mathcal{F}^2(m_q^2,q^2),\nonumber\\
\mathcal{M}_{B_4}&=&i^3 \int\frac{d^4 q}{(2\pi)^4}
	\Big[g_{\eta_{c} \mathcal{D}^{*} \mathcal{D}} \left(p_{2\mu}- p_{1\mu}\right)\Big] \left[e\Big(g^{\sigma\rho}(p_1^\chi-q^\chi)+g^{\chi\rho}p_1^\sigma+g^{\chi\sigma}q^\rho\Big)\epsilon_\chi^\gamma(p_3)\right] \Big[2f_{\mathcal{D}^*\mathcal{D}V}\varepsilon_{\kappa\tau\phi\xi}p_4^\kappa\epsilon^\tau(p_4)\big(p_2^\phi-q^\phi\big)\Big] \nonumber\\
	&\times&\frac{-g^{\alpha\sigma}+p_1^{\alpha} p_1^{\sigma} /m^2_1}{p^2_1-m^2_1} \frac{1}{p^2_2-m^2_2} \frac{-g^{\rho\xi}+q^{\rho} q^{\xi} /m^2_q}{q^2-m^2_q}\mathcal{F}^2(m_q^2,q^2),\nonumber\\
	\mathcal{M}_{B_5}&=&i^3 \int\frac{d^4 q}{(2\pi)^4}
	\Big[g_{\eta_{c} \mathcal{D}^{*} \mathcal{D}^{*}} \varepsilon_{\alpha\beta\nu\delta}p^{\delta} (p_{2\mu}- p_{1\mu})\Big] \Big[\frac{eg_{\mathcal{D}^*\mathcal{D}\gamma}}{4}\varepsilon^{\chi\lambda\sigma\rho}\epsilon_{\gamma}^{\eta}(p_3)\big(i p_{3\chi} g_{\lambda\eta}-i p_{3\lambda} g_{\chi\eta}\big)\big(i p_{1\sigma} g_{\rho\theta}-i p_{1\rho} g_{\sigma\theta}\big)\Big]\nonumber\\
	&\times &\Big[2f_{\mathcal{D}^*\mathcal{D}V}\varepsilon_{\kappa\tau\phi\xi}p_4^\kappa\epsilon^\tau(p_4)\big(q^\phi-p_2^\phi\big)] \frac{-g^{\alpha\rho}+p_1^{\alpha} p_1^{\rho} /m^2_1}{p^2_1-m^2_1} \frac{-g^{\beta\xi}+p_2^{\beta} p_2^{\xi} /m^2_2}{p^2_2-m^2_2} \frac{1}{q^2-m^2_q}\mathcal{F}^2(m_q^2,q^2),\nonumber\\
\mathcal{M}_{B_6}&=&i^3 \int\frac{d^4 q}{(2\pi)^4}
	\Big[ig_{\eta_{c} \mathcal{D}^{*} \mathcal{D}^{*}} \varepsilon_{\alpha\beta\nu\delta}p^{\delta} (p_{2\mu}- p_{1\mu})\Big] \left[e\Big(g^{\sigma\rho}(p_1^\chi-q^\chi)+g^{\chi\rho}p_1^\sigma+g^{\chi\sigma}q^\rho\Big)\epsilon_\chi^\gamma(p_3)\right] \Big[g_{\mathcal{D}^{*} \mathcal{D}^{*}V}g_{\tau \phi} \epsilon^{\tau}\left(p_{4}\right)\left(q^{\xi}-p^{\xi}_{2}\right)\nonumber\\
	&-&4f_{\mathcal{D}^{*} \mathcal{D}^{*}V} \epsilon^{\tau}\left(p_{4}\right)\left(p_4^{\xi}g_{\tau \phi}-p^{\phi}_{4}g_{\tau \xi}\right)\Big] \frac{-g^{\alpha\sigma}+p_1^{\alpha} p_1^{\sigma} /m^2_1}{p^2_1-m^2_1}\frac{-g^{\beta\xi}+p_2^{\beta} p_2^{\xi} /m^2_2}{p^2_2-m^2_2} \frac{-g^{\rho\phi}+q^{\rho} q^{\phi} /m^2_q}{q^2-m^2_q}\mathcal{F}^2(m_q^2,q^2).
\end{eqnarray}
\end{widetext}

\section{Discussion of ratios about the form-factor-power\label{Sec:Appetac1}}

\begin{table*}[htb]
	\renewcommand{\arraystretch}{1.5}
	\caption{Ratios of branching ratios for $\eta_c \to \mathcal{VV}$ and $\eta_c \to \gamma\mathcal{V}$ obtained with different form-factor powers.\label{Tab:4times}}
	\begin{tabular}{p{2.5cm}<\centering p{2.5 cm}<\centering p{2.5cm}<\centering p{2.5cm}<\centering p{2.5cm}<\centering}
		\toprule[0.8 pt]\toprule[0.8 pt]
		$N$ & $r^N_{12}$ & $r^N_{13}$ & $R^N_{12}$ & $R^N_{13}$\\
		\midrule[0.8pt]
		2 & 1.59 & 3.02$\sim$3.03  & 0.66$\sim$0.68  & 9.95$\sim$10.35 \\
     	3 & 1.62 & 3.40$\sim$3.62  & 0.61$\sim$0.67  & 9.01$\sim$10.53 \\
	    4 & 1.64 & 3.68$\sim$3.79  & 0.54$\sim$0.63  & 7.27$\sim$9.59  \\
		\bottomrule[0.8pt]  \bottomrule[0.8pt]
	\end{tabular}
\end{table*}

\begin{figure*}[htb]
	\centering
	\begin{tabular}{ccc}
		\includegraphics[width=8.4cm]{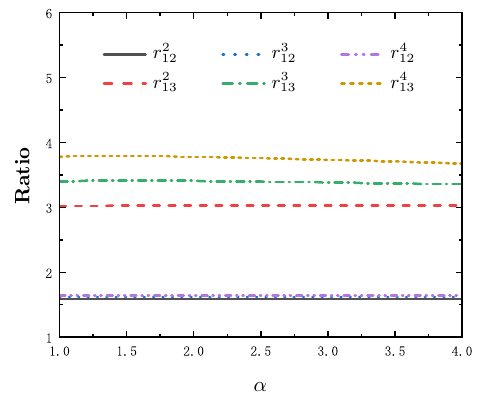}&
		$\hspace{0.5cm}$ &\includegraphics[width=8.4cm]{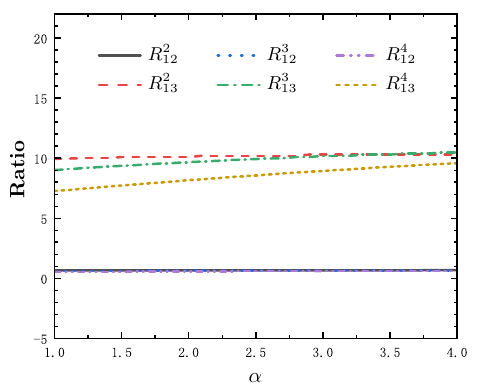}
	\end{tabular}
	\caption{(Color online.) Form factor dependence of the ratios of branching fractions for  $\eta_c \to \mathcal{VV}$ (left panel) and for  $\eta_c \to \gamma\mathcal{V}$(right panel). \label{Fig:r4}}
\end{figure*}

To further examine the parameter and form factor dependence of the ratios defined in Eqs.~\eqref{Eq:Ratio1} and \eqref{Eq:Ratio2}, we adopt form factors with different values of $N$ to evaluate the these ratios over a rather broad parameter range, $\alpha=1\sim 4 $. Here, we use the superscript $N$ in $r_{ij}^N$ and $R_{ij}^N$ to label distinct form factors, with $N$ being 2, 3, and 4. As shown in Fig.~\ref{Fig:r4}, the ratios $r_{12}^N$ and $R_{12}^N$ are nearly independent on the model parameter $\alpha$, and also exhibit weak dependence on the form factor. In particular, the ratio $r_{12}^N$ is around 1.6, while $R_{12}^N$ is approximately 0.6, as shown in Table~\ref{Tab:4times}. For $r_{13}^N$, this ratio remains fairly stable across this wide range of model parameter. As $N$ increases from 2 to 4, this ratio rises by roughly $20\%$. The $\alpha$ dependence of ratio $R_{12}^N$ becomes stronger as $N$ increase, and this ratio is estimated to be $9.95\sim 10.35$, $9.01\sim 10.53$ and $7.27\sim 9.59$ with $N=2$, $3$, and $4$, respectively. The different parameter and form factor dependence of these ratios can be attributed to the underlying meson-loop configurations. For the channels involving $\rho^0$ and $\omega$, the meson loops consist of identical charmed components, which leads to substantial cancellation of the form-factor dependence in the ratios. In contrast, loop contributions associated with the $\phi$ meson contain charmed-strange meson components, yielding a different meson-loop structure and a less complete cancellation of form-factor dependence.

\bibliographystyle{apsrev4-1}
\bibliography{etac1s}
\end{document}